\documentclass[runningheads]{llncs}
\usepackage[T1]{fontenc}
\usepackage{graphicx,verbatim}
\usepackage{amstext}
\usepackage{amsmath}
\usepackage{xcolor}
\usepackage{bbm}
\usepackage{booktabs}
\usepackage{float}
\usepackage{hyperref}
\usepackage{url}
\usepackage{algorithm}
\usepackage{algorithmicx}
\usepackage{algpseudocode}
\usepackage{soul}
\usepackage{subcaption}

\newcommand{\quotes}[1]{``#1''}
\begin{document}
%
\title{The Devil is in the Prompts: De-Identification Traces Enhance Memorization Risks in Synthetic Chest X-Ray Generation}
%



\author{Raman Dutt} 
\authorrunning{Raman Dutt et al.}
\institute{University of Edinburgh \\
    \email{raman.dutt@ed.ac.uk}}

\maketitle              
\begin{abstract}
Generative models, particularly text-to-image (T2I) diffusion models, play a crucial role in medical image analysis. However, these models are prone to training data memorization, posing significant risks to patient privacy. Synthetic chest X-ray generation is one of the most common applications in medical image analysis with the MIMIC-CXR dataset serving as the primary data repository for this task. This study presents the first systematic attempt to identify prompts and text tokens in MIMIC-CXR that contribute the most to training data memorization. Our analysis reveals two unexpected findings: \textbf{(1)} \emph{prompts containing traces of de-identification procedures (markers introduced to hide Protected Health Information) are the most memorized,} and \textbf{(2)} \emph{among all tokens, de-identification markers contribute the most towards memorization}. This highlights a broader issue with the standard anonymization practices and T2I synthesis with MIMIC-CXR. To exacerbate, existing inference-time memorization mitigation strategies are ineffective and fail to sufficiently reduce the model's reliance on memorized text tokens. On this front, we propose actionable strategies for different stakeholders to enhance privacy and improve the reliability of generative models in medical imaging. Finally, our results provide a foundation for future work on developing and benchmarking memorization mitigation techniques for synthetic chest X-ray generation using the MIMIC-CXR dataset. The anonymized code is available \href{https://anonymous.4open.science/r/diffusion_memorization-8011/README.md}{here}.

\end{abstract}

\section{Introduction}

High-quality data, often regarded as the "\textit{new gold}"\footnote{https://www.forbes.com/councils/forbestechcouncil/2023/03/27/how-to-make-use-of-the-new-gold-data/}, is vital in medical image analysis where large-scale datasets are scarce, hindering clinically viable AI development \cite{dutt2023parameter}. Diffusion models \cite{song2021denoising,ho2020denoising} have proven effective in producing novel, high-fidelity data.
In medical imaging, they address data scarcity while mitigating privacy, ethical, and legal challenges in data sharing \cite{yoon2020anonymization,murtaza2023synthetic}. Their efficacy is demonstrated in synthesizing radiographs \cite{chambon2022roentgen}, augmenting datasets \cite{saragih2024using,wang2024majority}, and enhancing downstream fairness \cite{ktena2024generative}, with ongoing advances promising further impact.

Generative models, despite their benefits, are prone to memorizing training data \cite{somepalli2023diffusion,somepalli2023understanding,wen2024detecting,dutt2024memcontrol,dutt2024capacity}, which threatens patient privacy. They may produce near-identical copies of training images, exposing sensitive details and enabling re-identification attacks that link synthetic outputs to real patients \cite{fernandez2023privacy}.

\textbf{The Unique Case of MIMIC-CXR:} Previous studies have linked memorization in diffusion models to the lexical structure of text prompts \cite{wen2024detecting}. Highly specific captions often act as keys into the model's memory, allowing the model to retrieve and replicate particular samples \cite{somepalli2023diffusion}, exhibiting memorization. MIMIC-CXR presents a distinct case as its text captions follow a structured phrase pattern, and multiple images often share identical captions due to similarities in clinical findings. For instance, in a filtered subset of 110K samples, 2337 instances share the caption \quotes{\textit{No acute cardiopulmonary abnormality.}}, indicating a normal finding. Furthermore, the publicly released version contains numerous traces of a specific marker (\quotes{\textcolor{red}{\textunderscore\textunderscore\textunderscore}}) used to de-identify the Protected Health Information (PHI) \footnote{https://www.hhs.gov/hipaa/for-professionals/privacy/index.html} which can further enhance caption specificity.

Given MIMIC-CXR's central role in developing T2I models for chest X-ray synthesis \cite{chambon2022roentgen,perez2024radedit,dutt2023parameter,dutt2024memcontrol}, it is crucial to investigate memorization at both the prompt and token levels to identify elements contributing most significantly to training data memorization. Similar analyses in natural image datasets \cite{webster2023duplication,schuhmann2022laion} have shaped benchmarks for detecting and mitigating memorization, underscoring the importance of conducting such a study for the medical imaging domain.     
 
To summarize, our core contributions are as follows:  
\textbf{(1)} We conduct the first systematic analysis to identify specific text prompts and tokens in MIMIC-CXR that contribute the most to memorization.  
\textbf{(2)} Our prompt-level (Sec \ref{sec:prompt_contribution}) and token-level (Sec. \ref{sec:token_contribution}) analysis uncovers a surprising yet concerning finding: \textcolor{red!60}{tokens introduced through standard de-identification procedures contribute the most to memorization.}  
\textbf{(3)} We release a comprehensive list of memorized prompts to facilitate future research on developing and benchmarking memorization mitigation techniques for synthetic chest X-ray generation using the MIMIC-CXR dataset.

\section{Related Work}

\textbf{Memorization in Generative Models: }Deep generative models have been shown to exhibit various forms of memorization, including training data extraction \cite{carlini2023extracting}, content replication \cite{somepalli2023diffusion}, and data copying \cite{somepalli2023understanding}. In the medical domain, \cite{akbar2023beware} found that diffusion models tend to memorize significantly more than GANs \cite{goodfellow}. Additionally, \cite{dar2023investigating} emphasized the need for robust mitigation strategies, highlighting the notable memorization in 3D Latent Diffusion Models (LDMs). \\\textbf{Mitigation Mechanisms: } Several mechanisms have been developed to mitigate memorization. \cite{somepalli2023understanding} introduced training and inference-time approaches, such as augmenting caption diversity. \cite{ren2024unveiling} presented a method that identifies memorized tokens by analyzing cross-attention scores, while \cite{wen2024detecting} devised an efficient procedure that leverages text-conditional noise for detection and mitigation. In medical image analysis, \cite{fernandez2023privacy} proposed a framework to remove samples that elevate memorization risk. Additionally, \cite{dutt2024memcontrol,dutt2024capacity} demonstrated that managing model capacity through Parameter-Efficient Fine-Tuning (PEFT) \cite{dutt2023parameter} can significantly reduce memorization. \\
Unlike prior studies that concentrate on mitigating memorization, our work underscores a fundamental flaw in data de-identification and employs established frameworks \cite{wen2024detecting,somepalli2023understanding} to demonstrate its connection to memorization.

\section{Preliminaries}

\subsection{Diffusion Models}

Diffusion models consist of two phases: forward and reverse diffusion. In the forward process, a data sample is gradually corrupted over $T$ steps by adding Gaussian noise according to a fixed Markov chain. At each step, the noise is injected as:
\begin{align} \label{eqn:forward}
    q\left(x_t \mid x_{t-1}\right)=\mathcal{N}(x_t ; \sqrt{1-\beta_t} x_{t-1}, \beta_t \mathbf{I}),
\end{align}
which leads to the closed-form expression
\begin{equation*} \label{eqn:step2}
    x_t = \sqrt{\overline{\alpha}_t}x_0 + \sqrt{1-\overline{\alpha_t}}\epsilon,
\end{equation*}

where $\overline{\alpha}_{t}=\prod_{i=1}^{t}(1-\beta_{t})$ \\
In the reverse process, one begins with a sample $x_T \sim \mathcal{N}(0,1)$ and iteratively denoises it to recover $x_0$. At each step, a learned noise estimator $\epsilon_{\theta}(x_t)$ predicts and subtracts the noise, updating the state as
\begin{equation*}
    x_{t-1} = \sqrt{\overline{\alpha}_{t-1}}\hat{x}_0^t + \sqrt{1 - \overline{\alpha}_{t-1}}\epsilon_{\theta}(x_t),
\end{equation*}
where $\hat{x}_0^t$ represents the intermediate estimate of $x_0$.

\subsection{Efficient Memorization Detection via Text-Conditional Noise} \label{sec:detection_framework}

A standard T2I stable-diffusion pipeline consists of a text encoder $T_E$, a variational autoencoder (VAE) $V_E$, and a noise predictor (U-Net). As noted in \cite{wen2024detecting}, for non-memorized prompts, the generated images are primarily influenced by the initial noise. In such cases, the model follows a denoising track influenced by both the initial noise and text-conditioning. However, for memorized prompts, the model overfits to a fixed denoising track, making the generated image largely independent of the initial noise. In this scenario, the model’s predictions become predominantly reliant on text-conditioning. 

\begin{figure}[htbp]
    \centering
    
    \subfloat[\textbf{ \textcolor{red}{(Memorized)} Prompt:} \textit{AP chest compared to \textcolor{black}{\textunderscore\textunderscore\textunderscore}: Previous mild pulmonary edema has resolved. There is no pneumonia ...
}
    ]{
        \includegraphics[width=0.8\textwidth]{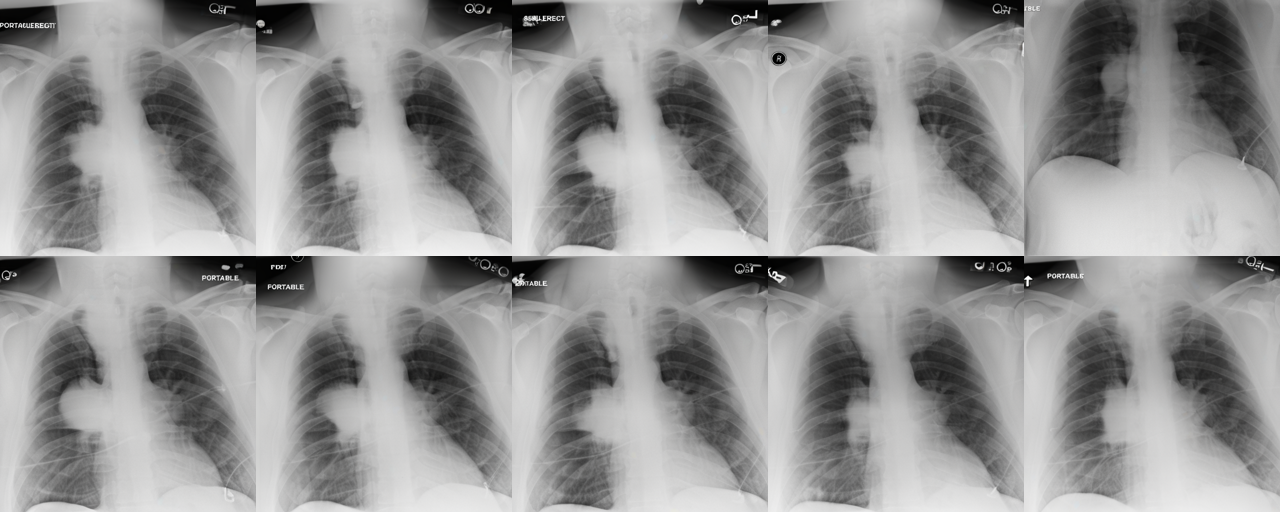}
    }
    
    \subfloat[\textbf{ \textcolor{cyan}{(Non-Memorized)} Prompt:} \textit{The right-sided chest tube, right-sided PICC line, and feeding tube are unchanged in position ... 
}
    ]{
        \includegraphics[width=0.8\textwidth]{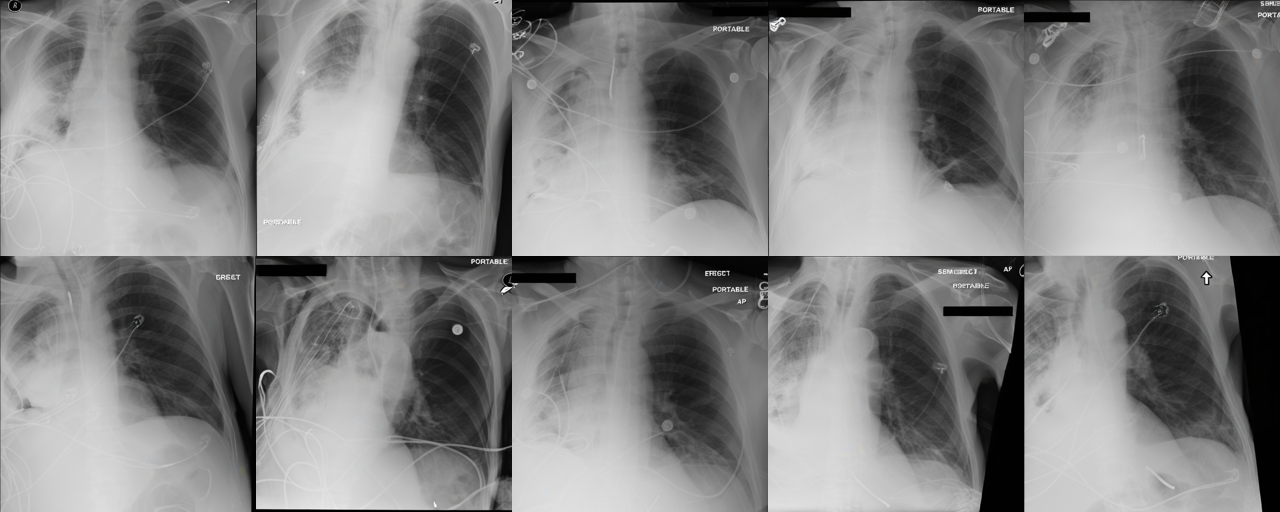}
    }
    
    \caption{Multiple generations for a single prompt across various initialization seeds. The top row shows a \textit{memorized} prompt, where images remain nearly identical regardless of the seed, indicating independence from initial noise. In contrast, the bottom row displays a \textit{non-memorized} prompt, with diverse outputs reflecting sensitivity to the initial noise, indicating no memorization.}
    \label{fig:multiple_generations}
\end{figure}

This phenomenon is demonstrated in Fig \ref{fig:multiple_generations}. For a prompt that has been identified as \quotes{\textit{memorized}}, the generations across multiple seeds show a striking resemblance to one another, indicating independence on the initial noise (controlled by the generation seed). On the contrary, multiple generations for a \quotes{\textit{non-memorized}} prompt, show differences with change in generation seed. 

Leveraging this insight, tracking the \emph{text-conditional noise} at each timestep emerges as a robust metric for detecting memorization \cite{wen2024detecting}. Given the noise predictor $\epsilon_{\theta}$ and $T$ timesteps, a prompt $p$ and an empty string $\emptyset$ with corresponding embeddings $e_p$ and $e_{\emptyset}$, the memorization detection metric $d_{mem}$ can be defined as:
\begin{equation*} 
    \boxed{ d_{mem} = \frac{1}{T} \sum_{t=1}^{T} \|\epsilon_{\theta}(x_t, e_p) - \epsilon_{\theta}(x_t, e_{\emptyset})\|_2. }
\end{equation*}


A higher value of $d_{mem}$ signifies a stronger memorization. This framework offers greater efficiency by providing a reliable memorization signal from the very first sampling step \cite{wen2024detecting}, making it well-suited for examining large datasets such as MIMIC-CXR. 


\section{Experiments}

\textbf{Experimental Setup.} A reliable memorization signal necessitates an in-domain latent diffusion model capable of generating high-quality chest X-rays. For this task, we employ the off-the-shelf \textit{RadEdit} model \cite{perez2024radedit}, which integrates a biomedical text encoder \cite{bannur2023learning} and the VAE from SDXL \cite{podell2023sdxl}. This model is particularly well-suited to our setup as it includes the MIMIC-CXR dataset in its training corpus. For detecting memorization in prompts, we employ the framework from \cite{wen2024detecting} (Sec \ref{sec:detection_framework}) due to its reliability and efficiency.

\subsection{Detecting Memorized Prompts in MIMIC-CXR} \label{sec:prompt_contribution}

\begin{figure}[htb] 
    \centering
    \includegraphics[width=0.9\textwidth]{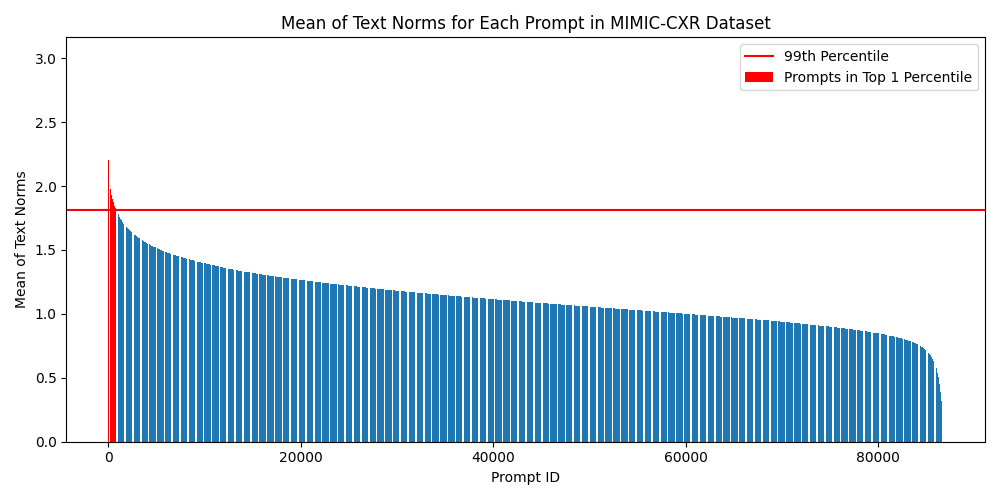} 
    \caption{Visualizing the distribution of text-conditional norms for unique prompts in the MIMIC-CXR dataset (largest to smallest). Prompts in the top 1 percentile, exhibiting the highest norms, are highlighted in red. Prompts exhibiting high norms indicate they are potentially memorized.}
    \label{fig:memorization_framework}
\end{figure}

\boxed{\textbf{Setup:}} To identify all memorized prompts in the MIMIC-CXR dataset, we begin by extracting the subset of all unique prompts. Using a text-to-image pipeline comprising a pre-trained denoising U-Net ($\epsilon$), a text-encoder ($T_E$) and a VAE ($V_E$), we track and store the text-conditional noise for each unique prompt at every denoising timestep. Finally, we compute the average text-conditional noise across all timesteps to quantify memorization. This gives us a memorization score ($d_{mem}$) for each unique prompt in the dataset.
\\ \boxed{\textbf{Results:}} Figure \ref{fig:memorization_framework} illustrates the distribution of the memorization scores for all unique prompts, sorted in descending order for visual clarity. The distribution follows a \textit{heavy-tailed} pattern, with a small subset of prompts (on the left) exhibiting significantly higher norms, indicating a stronger contribution to memorization. The prompts corresponding to the top 1 percentile of norm values, highlighted in red and referred to as \quotes{\textcolor{red}{\textit{memorized prompts}}} hereafter, represent the most extreme cases indicating the highest contribution towards memorization. The gradual decline in norm values across the remaining prompts suggests a varying degree of influence on memorization, with the majority exhibiting relatively lower norms. This variability underscores the need for further investigation into prompts with the highest memorization scores, as they may reveal underlying patterns that contribute to memorization risks. We conduct further analysis in section \ref{sec:token_contribution}.

\subsection{Examining Individual Token Contribution: Traces of De-Identification Enhance Memorization} \label{sec:token_contribution}

\begin{figure}[htb]
    \centering
    
    \subfloat[\textbf{Prompt:} \textit{AP chest compared to \textcolor{red}{\textunderscore\textunderscore\textunderscore}: Previous mild pulmonary edema has resolved.  There is no pneumonia. Several small lung nodules and the large right paratracheal mediastinal mass are manifestations of lung cancer. Heart size normal.  No appreciable pleural effusion.}
    ]{
        \includegraphics[width=0.9\textwidth]{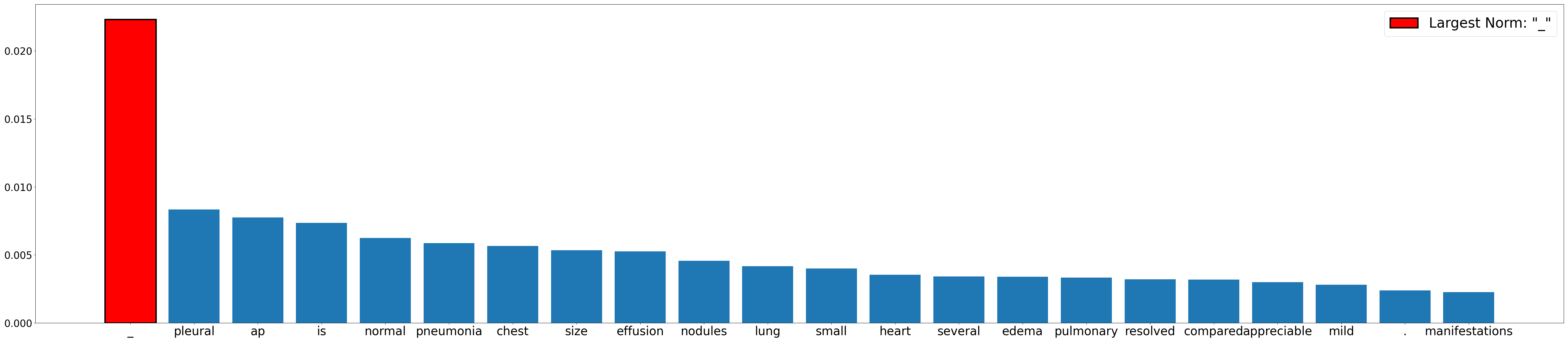}
    }
    \caption{Figure illustrating the text-conditional norm for each token in a memorized prompt. We only plot the tokens with the top 25 norm values for visual clarity. Amongst all tokens, the PHI de-identification token (\quotes{\textcolor{red}{\textunderscore\textunderscore\textunderscore}}) holds the most significant contribution towards memorization. This behaviour is replicated across all memorized prompts.}
    \label{fig:token_contribution}
\end{figure}

\boxed{\textbf{Token-Level Analysis:}} Building on the \textit{prompt-level} analysis in Section \ref{sec:prompt_contribution}, we extend our investigation to the \textit{token-level}. Specifically, we focus on the set of \textit{memorized prompts} and analyze the contribution of individual tokens toward memorization. \\
\boxed{\textbf{Results:}} Our findings consistently show that within memorized prompts, the de-identification marker is the token contributing most significantly to memorization, as illustrated in Figures \ref{fig:token_contribution}. We hypothesize two key reasons for this phenomenon: \textbf{(1)} The de-identification marker is a distinct and unique token, differing from all other tokens in the MIMIC-CXR text corpus. \textbf{(2)} It appears frequently across the dataset, occurring in 21,373 unique prompts. This high frequency allows the model to learn spurious correlations, leading to the memorization of specific samples. 
This finding is particularly concerning as de-identification is a standard practice before publicly releasing medical datasets. Our results highlight the need to reassess current de-identification methodologies to prevent unintended memorization in generative models.

\subsection{Existing Intervention Methods are Ineffective}

\begin{figure}[htb] 
    \centering
    \includegraphics[width=0.9\textwidth]{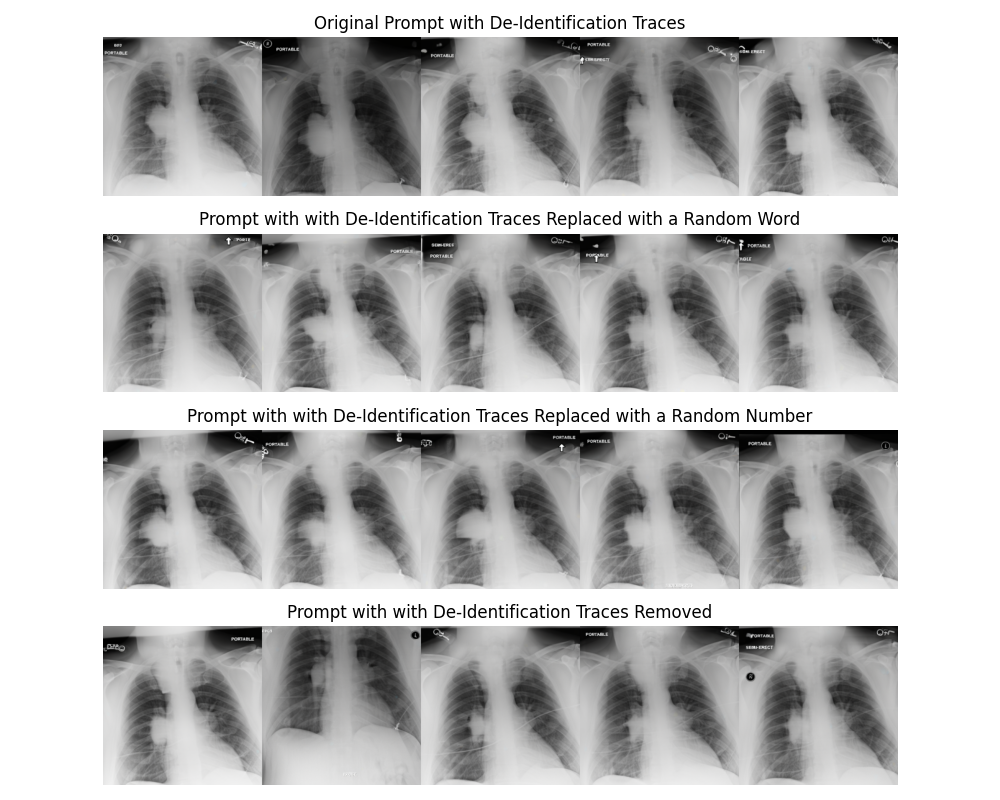} 
    \caption{Figure depicting multiple generations for the same prompt and different mitigation strategies. The visual similarity across different generations and mitigation methods indicates their ineffectiveness.}
    \label{fig:mitigation_strategies}
\end{figure}

In this section, we investigate whether applying memorization mitigation strategies to de-identification traces can effectively reduce memorization. Specifically, we evaluate different inference-time mitigation techniques \cite{somepalli2023understanding}: \textbf{(1)} \textit{\textbf{Random Word Addition (RWA)}}, where de-identification markers are replaced with random words; \textbf{(2)} \textit{\textbf{Random Number Addition (RNA)}}, where markers are substituted with random numbers; and \textbf{(3)} the \textbf{complete removal} of de-identification markers from the prompt. \\ \boxed{\textbf{Results:}} We assess memorization by analyzing multiple generations across different initialization seeds for the same prompt. Memorization is qualitatively indicated by the similarity among generated images. For a quantitative evaluation, we compute the mean L2 distance between 50 generated samples using the same prompt, where a lower L2 distance signifies stronger memorization.
Across all mitigation strategies, we observe that the model continues to generate visually similar images. Simply replacing de-identification markers with a random word or number, or even removing them entirely, remains ineffective. Quantitative analysis reinforces this observation. The average L2 distance over 50 generations remains nearly unchanged after applying mitigation strategies: 0.38 for the original prompt versus 0.45, 0.43, and 0.42 with mitigation strategies applied. These findings indicate a deeper underlying issue that must be addressed at the training level.

\section{Discussion and Conclusion}
This section examines potential factors through which de-identification practices may inadvertently heighten the risks of memorization and compromise privacy preservation. We also offer recommendations for medical AI researchers involved in dataset curation, pre-processing, and the training of T2I models with a focus on mitigating memorization. \\ \boxed{\textbf{Why Do de-identification Markers Lead to Memorization?}} The text corpus in MIMIC-CXR exhibits a distinct lexical structure, notably marked by the frequent occurrence of the de-identification token (\quotes{\textcolor{red}{\textunderscore\textunderscore\textunderscore}}). Introduced during the de-identification process, this token offers no substantive information for text-to-image generation. Instead, it creates a spurious correlation with the corresponding images. As a result, such highly specific tokens can serve as retrieval keys, allowing for the extraction of particular data points that appear as repeated, replicated generations, indicating memorization, as shown in \cite{somepalli2023understanding}. \\ \boxed{\textbf{Recommendations for Enhancing Privacy Preservation: }}  
We propose several actionable strategies for different stakeholders. \\\textbf{Dataset curators} should refrain from using a uniform de-identification marker across the entire dataset. By employing a rule-based de-identification approach as in \cite{johnson2016mimic}, curators can randomize the marker symbols. This method not only enhances the diversity of captions that can mitigate memorization \cite{somepalli2023understanding} but also helps to minimize the risk of establishing spurious correlations between specific tokens and images. \\
\textbf{Model developers} tasked with training T2I models should invest additional effort in pre-processing dataset captions. For example, recaptioning datasets to eliminate redundant tokens can enhance both the quality and diversity of the captions. Additionally, employing an in-domain vision-language model (VLM) \cite{llavamed} can refine the language and augment the information density of the captions. This strategy is expected to improve caption diversity and boost generative performance \cite{segalis2023pictureworththousandwords}. \\\\ In summary, our work tackles the challenges of memorization and privacy preservation. By focusing on MIMIC-CXR, the most widely used dataset for T2I generation of chest X-rays, we reveal a critical flaw in the conventional de-identification procedure employed in medical datasets, establishing a clear connection to memorization. Moreover, we demonstrate that removing memorization from trained models is a complex task, with standard mitigation techniques falling short. To address this issue at its source, we offer targeted recommendations for various stakeholders. Finally, we release a list of memorized prompts to support future benchmarking and the development of more effective mitigation strategies.

\newpage
\clearpage

\bibliographystyle{splncs04}
\bibliography{mybibliography}

\end{document}